# Magnetically tunable zero-index metamaterials

Yucong Yang[1,2], Yueyang Liu[3], Jun Qin[1,2], Songgang Cai[1,2], Jiejun Su[1,2], Peiheng Zhou[1,2], Longjiang Deng[1,2*], Yang Li[3*] and Lei Bi[1,2*]

[1]National Engineering Research Centre of Electromagnetic Radiation Control Materials, University of Electronic Science and Technology of China, Chengdu 610054, China

[2]State Key Laboratory of Electronic Thin-Films and Integrated Devices, University of Electronic Science and Technology of China, Chengdu 610054, China

[3]State Key Laboratory of Precision Measurement Technology and Instrument, Department of Precision Instrument, Tsinghua University, Beijing 100084, China

Corresponding author

denglj@uestc.edu.cn, yli9003@mail.tsinghua.edu.cn, bilei@uestc.edu.cn

## Abstract

Zero-index metamaterials (ZIMs) feature a uniform electromagnetic mode over a large area in arbitrary shapes, enabling many applications including high-transmission supercouplers with arbitrary shapes, direction-independent phase matching for nonlinear optics, and collective emission of many quantum emitters. However, most ZIMs reported till date are passive, with no method for the dynamic modulation of their electromagnetic properties. Here, we design and fabricate a magnetically tunable ZIM consisting of yttrium iron garnet (YIG) pillars sandwiched between two copper clad laminates in the microwave regime. By harnessing the Cotton-Mouton effect of YIG, the metamaterial was successfully toggled between gapless and bandgap states, leading to a "phase transition" between a zero-index phase and a single negative phase of the metamaterial. Using an S-shaped ZIM supercoupler, we experimentally demonstrated a tunable supercoupling state with a low intrinsic loss of 0.95 dB and a high extinction ratio of up to 30.63 dB at 9 GHz. Our work enables dynamic modulation of the electromagnetic characteristics of ZIMs, enabling various applications in tunable linear, nonlinear, quantum and nonreciprocal electromagnetic devices.

## Introduction

Zero-index materials are materials or composite structures that exhibit an effective refractive index of zero at a given frequency[1-4], resulting in an infinite spatial wavelength. This effect can be leveraged to overcome the limitations imposed by the finite spatial wavelength of electromagnetic waves, thereby enabling various novel physical phenomena and applications in linear[5-7], nonlinear[8-10] and quantum electromagnetic systems[11,12]. Recently, zero-index materials, such as indium tin oxide[10,13,14], waveguides at cut-off frequencies[15-17], fishnet metamaterials[18], and doped $\varepsilon$-near-zero (ENZ) media[19], have garnered significant research interest[2]. However, the aforementioned materials and artificial structures exhibit large ohmic losses because of their metallic components[20]. In contrast, zero-index metamaterials (ZIMs) based on all-dielectric photonic crystals exhibit zero ohmic loss, enabling the realization of ZIMs over a large area of arbitrary shapes. A photonic crystal-based ZIM was first realized in the microwave regime based on $Al_2O_3$ pillars embedded in a parallel metal waveguide[21]. Subsequently, photonic-crystal-based ZIM ranging from acoustic[22,23] to photonic regimes[24,25] have been reported, demonstrating fascinating physical phenomena and applications, such as supercoupling[1,26], leaky-wave antennas[27], cloaking[21,28], superradiance[12,25], and direction-independent phase matching for nonlinear optics[9].

Despite this progress, most of the ZIMs reported thus far are passive, with constant post-fabrication electromagnetic properties, limiting their applications in passive devices. Active ZIMs, whose magnetic ($\mu_{eff}$) and electric ($\varepsilon_{eff}$) properties can be tuned using external stimuli, may allow the dynamic tuning of delicate photonic band structures, thereby inducing "phase transitions" in metamaterials. In turn, this unique mechanism is expected to enable energy-efficient modulation of electromagnetic wave propagation with low insertion loss, high extinction ratio, and compact device footprint—which are all essential factors in microwave and optical communication applications[4,19,29,30].

In this study, we design and experimentally investigate a magnetically tunable ZIM consisting of an array of gyromagnetic pillars embedded within a parallel-plate copper waveguide. Under an applied magnetic field of 430 Oe, the photonic band structure of the proposed ZIM changes from a zero-index state to a photonic bandgap state, corresponding to a transition from a "zero-index phase" to a "single negative phase". Based on this property, we propose a magnetic field-induced on-off switch of the supercoupling state in an S-shaped ZIM waveguide, resulting in a low intrinsic loss of 0.95 dB and a high extinction ratio of 30.63 dB at 9 GHz. We also demonstrate a magnetic field-controlled switch to effect transitions between a zero-index state and a nontrivial topological boundary state in the magnetic ZIM. These results demonstrate the potential of applying active ZIMs in electromagnetic wave modulators and nonreciprocal devices.

## Results

First, we designed a Dirac-like cone-based zero-index metamaterial (DCZIM) consisting of a square array of dielectric pillars embedded in a parallel-plate copper waveguide[21]. At the Γ point, the accidental degeneracy of two linear dispersion bands and a quadratic dispersion band formed a Dirac-like cone dispersion, corresponding to an impedance-matched zero effective index[21]. In contrast to conventional passive ZIMs, we achieved active modulation by fabricating a square lattice of pillars constructed using a magnetic dielectric material, yttrium iron garnet (YIG). By applying a magnetic field to the YIG pillars along the direction perpendicular to the wave vector of the transverse magnetic (TM) mode, an effective permeability modulation that is quadratically proportional to the YIG magnetization was observed owing to the Cotton-Mouton effect (see Supplementary Information S1 for further details). In turn, this effect enabled the modulation of the photonic band structure and the effective index of metamaterials.

We implemented the proposed design using the structure depicted in Figure 1. Figure 1a illustrates the experimentally fabricated metamaterial consisting of a square lattice of gyromagnetic YIG pillars with a 3.53-mm radius and a 17.9-mm lattice constant. The dielectric constant[31] and permeability of the YIG material were characterized (see the Supplementary Information S2 for further details). The YIG pillars were placed in a waveguide consisting of two parallel copper clad laminates separated by 4 mm. A magnetic field was applied to each YIG pillar by placing a neodymium iron boron (NdFeB) permanent magnet under each pillar and behind the copper back plate. A uniform magnetic field along the $z$-direction was observed, whose intensity reached 430 Oe in the middle of the two copper clad laminates. This was sufficient to saturate the YIG pillars (see Supplementary Information S3 for further details).

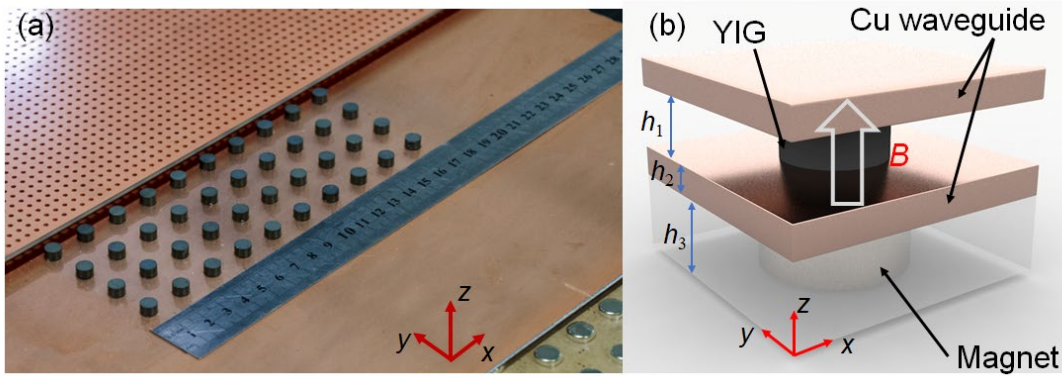

**Figure 1. Schematic diagram of the active DCZIM structure.**

**(a)** The structure of an active DCZIM based on a gyromagnetic photonic crystal. **(b)** Schematic diagram of a unit cell. The YIG pillars were placed in a parallel-plate copper-clad waveguide with height, $h_1$ = 4 mm. The thickness of the waveguide plates was $h_2$ = 2 mm. Permanent magnets with 5-mm diameter and height $h_3$ = 5 mm were placed in an acrylic matrix underneath the waveguide and were aligned with the YIG pillars.

To characterize the properties of DCZIM, we first calculated and then experimentally measured the photonic band structure in the plane of the YIG array. The gray dotted line in Figure 2a represents the calculated band structure for the TM modes (the electric field is polarized in the *z* direction) of this photonic crystal, as observed based on a simulation using COMSOL MULTIPHYSICS. We observed a clear Dirac-like cone dispersion at 9 GHz. When a magnetic field of +430 Oe was applied along the *z* direction, the off-diagonal component induced the Cotton-Mouton effect in YIG, changing the frequencies of the three photonic bands forming the Dirac-like cone. Owing to time reversal symmetry (TRS) breaking, the photonic crystal transitioned from $C_4^z$ symmetry to $C_2^z$ symmetry (see Supplementary Information S4 for further details). As a result, the degeneracy was broken, resulting in two bandgaps, as indicated by the gray dotted lines in the right panel of Figure 2a.

The modes supported by this structure were analyzed by considering those supported by a square array of two-dimensional YIG pillars. As depicted in Figure 2b–d, this structure supported three modes for TM polarization near the 9 GHz frequency—a monopole mode, a transverse magnetic dipole mode, and a longitudinal magnetic dipole mode. When a magnetic field was applied, as illustrated in the right part of the panel, these three modes were displaced to different frequencies (8.95 GHz, 9.52 GHz, 11.04 GHz, respectively). Because of the broken TRS, all three modes required to be rotated through 180 ° to coincide with each other.

We further calculated the Chern number of each photonic band from low frequency to high frequency near the Dirac point to be 0, 1, and -1, respectively. Based on the Chern number of each band, we calculated the Chern numbers of the bandgaps, 1 and 2 (right panel of Figure 2a), to be $\Delta C_{\Gamma 1-2}=|C_{\Gamma 1}|=0$ and $\Delta C_{\Gamma 2-3}=|C_{\Gamma 1}+C_{\Gamma 2}|=1$, respectively. Based on the Chern numbers of the bandgaps, 1 and 2, we can determine their topological nature—bandgaps 1 and 2 were topologically trivial and nontrivial, respectively.

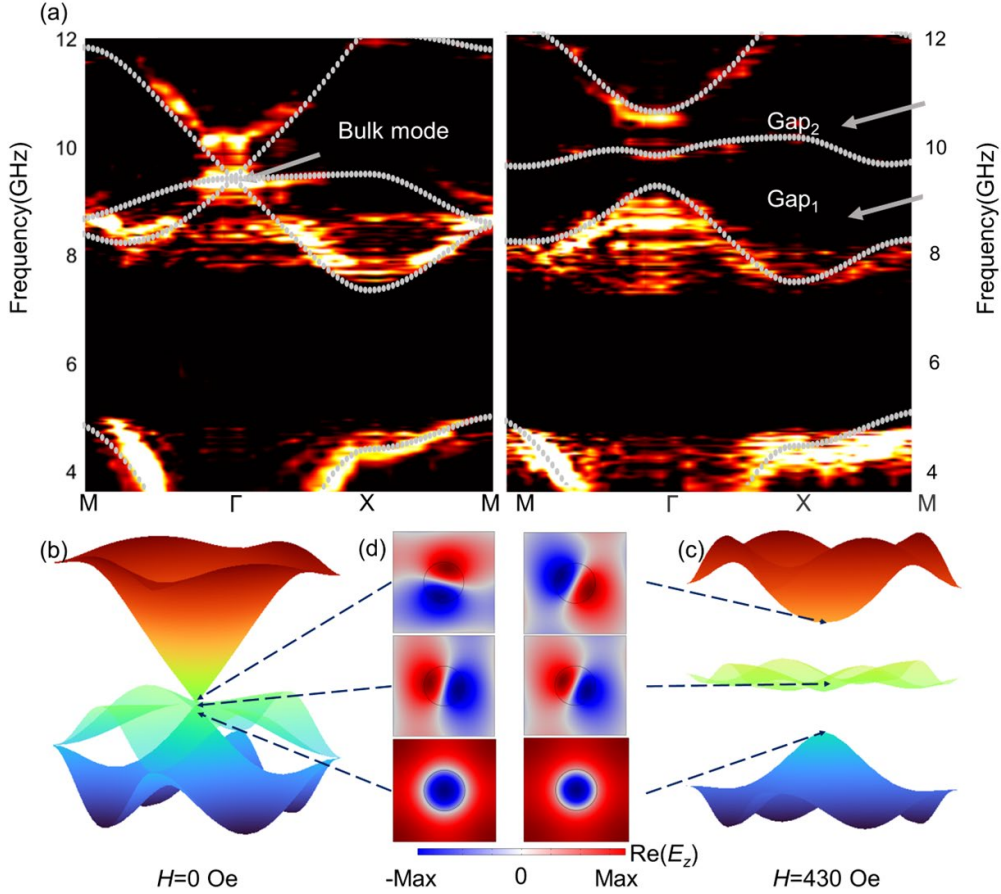

**Figure 2. Theoretical and experimental demonstration of an active ZIM. (a).** Measured and calculated (grey dots) photonic bands of the active ZIM using Fourier transform field scan (FTFS) of the TM modes corresponding to applied magnetic fields of 0 (left panel) and 430 Oe (right panel). **(b)–(c).** Simulated three-dimensional dispersion surfaces near the Dirac-point frequency, depicting the relationship between the frequency and the wave vectors ($k_x$ and $k_y$) **(d).** COMSOL-computed Re($E_z$) on the cross-section of a ZIM unit cell at the frequencies indicated by dashed arrows, depicting an electric monopole mode, a transverse magnetic dipole mode, and a longitudinal magnetic dipole. The black circles indicate the boundaries of the YIG pillars.

We experimentally characterized this metamaterial using the setup proposed by Zhou et al.[31] The sizes of all the samples were designed to be 10 × 10 periods, as illustrated in Figure 1a. The parallel-plate waveguide consisting of two copper clad laminates separated by 4 mm supported only the fundamental TEM mode in a parallel-plate waveguide below 37.5 GHz. To facilitate the excitation and measurement of electromagnetic fields inside the waveguide, a square array of holes, with a lattice constant of 5 mm, was drilled through the top surface of the copper-clad laminate, enabling the insertion of a probe for field measurement. The thickness of the copper-clad laminates was taken to be 2 mm to tune the uniform distribution of the magnetic flux.

First, we measured the photonic band structure of the metamaterial under an applied magnetic field of 0, as illustrated using the intensity plot in the left panel of Figure 2a. The band structure of the TM bulk states was obtained by applying two-dimensional discrete Fourier transform (2D-DFT) to the measured complex field distribution over the metamaterial (see Supplementary Information S5 for further details). The measured photonic band structure exhibited good agreement with the simulation results (represented by gray dots). Both the measured and computed band structures exhibited a bandgap between 5 GHz and 7 GHz as well as bulk modes between 7 GHz and 12 GHz, indicating a Dirac-like cone dispersion near 9 GHz. The nondegeneracy of the photonic bands was also recorded after applying a 430 Oe magnetic field along $z$-direction, as depicted in the right panel in Figure 2a. The bandgaps were experimentally measured to be at 8.95–9.60 GHz and 10.24–11.04 GHz, corroborating the simulation results.

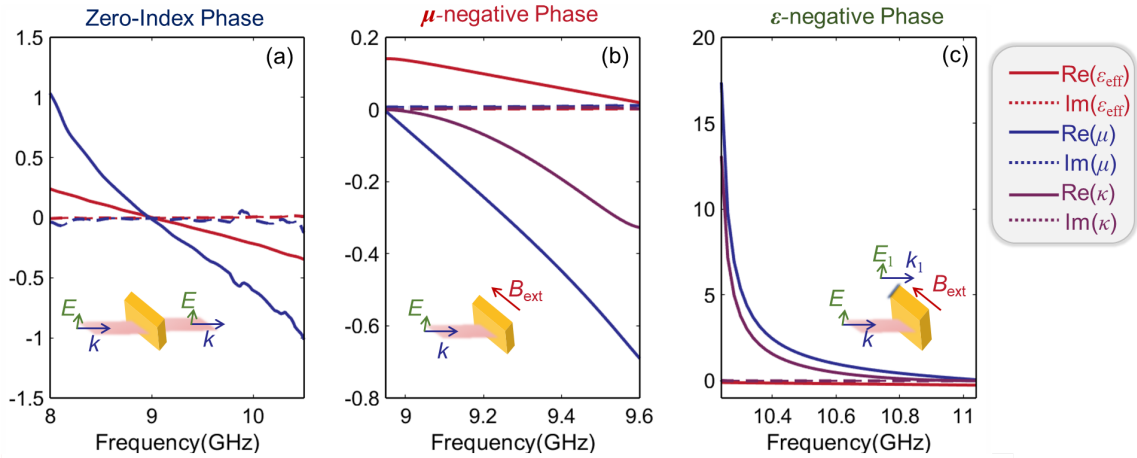

**Figure 3. Magnetic field-induced phase transition of active ZIM.** Real and imaginary parts of the effective permittivity ($\varepsilon_{\text{eff}}$) and permeability tensor elements ($\mu$ and $\kappa$) **(a)** under an applied magnetic field of 0, and **(b)** under an applied magnetic field of 430 Oe in the bandgap frequency range of 8.95–9.60 GHz and **(c)** the bandgap frequency range of 10.24–11.04 GHz

The magnetic field-induced band structure and transmittance modulation can be regarded to be a phase transition process from a material perspective. This phenomenon can be observed by retrieving the effective permittivity and permeability tensor elements of the metamaterial in the presence and absence of an applied magnetic field, as illustrated in Figure 3. We used the boundary effective medium approach (BEMA) to calculate the effective constitutive parameters, as proposed in a previous study on ZIM[32]. In Figure 3a–c, $\varepsilon_{eff}$ denotes the effective permittivity, $\mu$ denotes the diagonal of the effective permeability tensor, and $\kappa$ denotes the off-diagonal component of the effective permeability tensor, $\begin{pmatrix} \mu & i\kappa & 0 \\ -i\kappa & \mu & 0 \\ 0 & 0 & 1 \end{pmatrix}$.

Figure 3a depicts the results corresponding to an applied magnetic field of 0. The real parts of $\varepsilon_{eff}$ and $\mu$ cross zero simultaneously and linearly at 9 GHz, exhibiting ε-and-μ-near-zero (EMNZ) behavior corresponding to the zero-index phase. The imaginary parts of $\varepsilon_{eff}$ and $\mu$ were both close to 0. This low loss was attributed to the small loss tangent (0.0002) of the YIG material in this frequency range. When a magnetic field of 430 Oe was applied, phase transitions occurred from the zero-index phase to the μ-negative (MNG) or the ε–negative (ENG) phase, as depicted in Figures 3b and 3c, respectively. In the bandgap, 8.95 GHz–9.6 GHz (Figure 3b), $\varepsilon_{eff}$ is positive, whereas $\mu_{eff}$ ($\mu_{eff} = \dfrac{\mu^2 - \kappa^2}{\mu}$)[33] is negative, which corresponds to the MNG phase. The impedance at 9 GHz was tuned from 1.84 to 0.62i after applying the magnetic field, leading to a total reflection of the incident EM wave owing to impedance mismatch. In the bandgap, 10.24–11.04 GHz (Figure 3c), $\varepsilon_{eff}$ is negative, whereas $\mu_{eff}$ is positive, which corresponds to the ENG phase. Corresponding to both frequency ranges, the real parts of $\varepsilon_{eff}$ and $\mu_{eff}$ decreased as the frequency increased, exhibiting anomalous dispersion. The incident EM wave was still reflected due to impedance mismatch. Additionally, as depicted in the inset of Figure 3c in the ENG frequency regime, a nontrivial topological boundary state was observed at the metamaterial edge because of the difference between the Chern numbers of the upper and lower structures.

The complex phase transition phenomena discussed above were attributed to the location of the ZIM at the origin of the metamaterial phase diagram, which allowed it to reach all quadrants of the phase diagram via appropriate tuning of the constitutive parameters[33,34]. Further discussion on the attainment of other phases based on the phase diagram is depicted in Figures S5–7 of Supplementary Information S6–7. Such unique properties of an active ZIM are indicative of its potential with respect to the modulation of the propagation of EM waves.

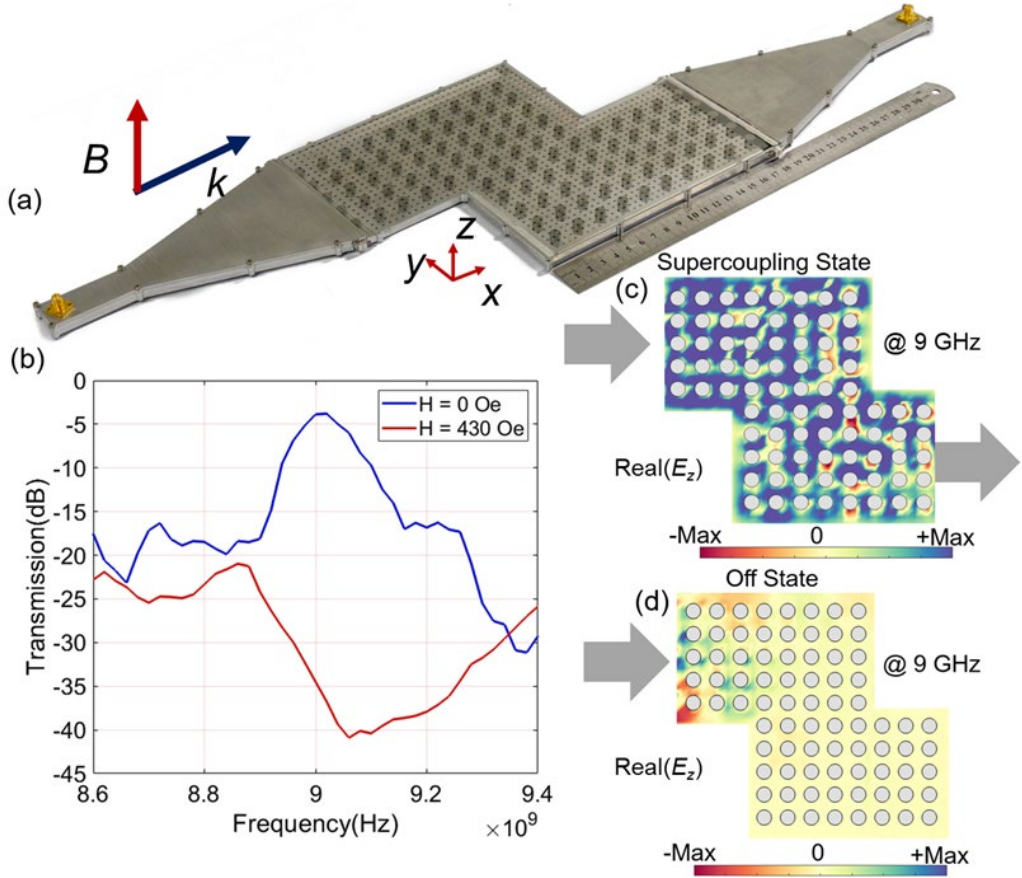

**Figure 4 Structure and characterization of a microwave switch based on active DCZIM. (a)** Photograph of the microwave switch sample. **(b)** The measured transmissions in the absence and presence of an applied magnetic field of 430 Oe. **(c)** The real part of the $E_z$ distribution observed at 9 GHz inside the metamaterial (each pillar is indicated in gray) in the absence of a magnetic field. **(d)** The real part of the $E_z$ distribution observed at 9 GHz inside the metamaterial (each pillar is indicated in gray) in the presence of a 430 Oe magnetic field.

To showcase a microwave switch based on the phase transition effect of active ZIM, we fabricated a ZIM waveguide switch by leveraging the contingency of the supercoupling state on the applied magnetic field, as illustrated in Figure 4a. First, a ZIM waveguide comprising top and bottom metal plates and 100 YIG pillars was fabricated, forming two sharp 90-degree bends, to verify the supercoupling effect experimentally (see Supplementary Information S8 for further details). The waveguide was coupled to the coaxial line via SMA connectors. Linear tapered sections were used to sustain the $TE_{10}$ mode and induce its gradual evolution into the TM mode of the metamaterial during propagation from the source to the waveguide. Perfect magnetic conductor (PMC) boundary conditions were realized using aluminum alloy walls at a distance of $\lambda/4$ from the metamaterial as the lateral boundaries[26] (see Supplementary Information S8 for further details). Using the device depicted in Figure 4a, we successfully switched between the bulk state (supercoupling state) and the photonic bandgap state (off state) by applying appropriate magnetic fields to the YIG pillars.

Figure 4b depicts the measured transmission spectra corresponding to both states. A large transmission contrast was observed over 8.9–9.4 GHz, which was consistent with the bandgap frequencies calculated via numerical simulation in Figure 2a. The difference in device transmission under zero and 430 Oe magnetic fields was larger than 30 dB at approximately 9 GHz and the device insertion loss was 3.75 dB. Considering that the coupling loss induced by the SMA connectors was 2.8 dB, the intrinsic loss of the ZIM waveguide was as low as 0.95 dB, primarily induced by the absorption of YIG materials.

We verified the supercoupling behavior in the absence of an applied magnetic field by measuring the $E_z$ component of the electric field at each point of the metamaterial, as illustrated in Figure 4c. At 9 GHz, the electric field tunneled through the metamaterial with almost no phase change in the form of a bulk mode, verifying supercoupling behavior. In contrast, in the absence of YIG pillars in the waveguide, the wave was reflected back to the incident port, as described in Supplementary Information S9. This confirmed that the supercoupling behavior was induced by the DCZIM. As depicted in Figure 4d, in the presence of an applied magnetic field, the electromagnetic wave decayed exponentially in the metamaterial owing to the photonic bandgap 1 depicted in Figure 2a, leading to a high extinction ratio. We also constructed a switch between the supercoupling state and the topological one-way transmission state by probing at the upper bandgap frequency of 10.6 GHz (see Supplementary Information S10 for further details). These unique properties are indicative of the potential of active ZIMs in novel active electromagnetic devices.

## Discussion

In this study, we proposed and experimentally operated a magnetically tunable ZIM. The metamaterial was operated by leveraging the Cotton-Mouton effect of the constitutive YIG pillars under applied magnetic fields, which alter the symmetry and bandgap opening of the Dirac-like cone-based ZIM. From a material perspective, the proposed metamaterial exhibited a phase transition from the zero-index phase to a single negative phase, leading to an effective index change from 0 to 0.09i at 9 GHz. Based on this property, we constructed and verified the function of a microwave switch by manipulating the supercoupling effect, thereby reducing the intrinsic loss to 0.95 dB and achieving a high extinction ratio of 30.63 dB at 9 GHz.

We believe that this study introduces a new approach to active ZIMs, particularly with respect to the development of efficient active electromagnetic and nonreciprocal devices. By appropriately engineering the slots in parallel-plate copper waveguides[27], continuous beam steering over the broadside can be achieved by varying the applied magnetic field. Moreover, our design can be extended to the optical regime by embedding YIG pillars in a polymer matrix with gold films cladded[24], enabling the dynamic modulation of optical DCZIM. Further, such a magnetically tunable DCZIM can be used to modulate the four-wave mixing process in DCZIM by manipulating the zero-index phase-matching condition[9]. Additionally, we were able to modulate DCZIM-based large-area single-mode photonic crystal surface-emitting lasers with high output power[35]. Finally, we also successfully modulated the extended superradiance by tuning the effective index of the DCZIM, in which many quantum emitters were embedded[12].

## Methods

**Numerical simulation.** The permittivity of the YIG used during numerical simulation was taken from that published by Zhou et al.[36] The band structure was calculated using the COMSOL software. The results were obtained by calculating the modes with periodic boundary conditions along the $x$ and $y$ directions. The TM polarization mode was selected by considering the electric field to be an out-of-plane vector. Under magnetic fields, the DCZIM was simulated to obtain frequency-dependent electromagnetic field profiles. The transitions between the different metamaterial phases in the presence and absence of magnetic fields were simulated by considering the off-diagonal elements of the YIG permeability tensor. Home-made MATLAB codes based on BEMA and 2D-DFT were used to calculate the effective permittivity, permeability tensor, and photonic band structure.

**Device Fabrication.** The active DCZIM consisted of a square array of YIG pillars with a radius of 3.53 mm, a height of 4 mm, and a lattice constant of 17.9 mm (as indicated in Figure 1a). The waveguide was formed using two 500 mm × 500 mm copper-clad laminates. During the measurement of the nontrivial boundary state, a copper bar with a length of 40 cm and a width of 2 cm was placed at the edge of the metamaterial to form its boundary.

The YIG pillars were fabricated from YIG bulk crystals using an ultrahigh-accuracy computer numerical control (CNC) machine (Mazak VARIAXIS i-700) with a dimensional accuracy exceeding 0.05 mm. As depicted in Figure 1a, an external magnetic field was introduced to maintain the saturation magnetization of the garnet. To affix the YIG pillars to the copper clad laminate substrate tightly, double-sided tape with a radius of 3.53 mm was applied to one side of each pillar. The position of each YIG pillar was precisely defined by placing the other side in an acrylic mold with 10 × 10 holes with a radius of 3.54 mm and a lattice constant of 17.9 mm. The copper clad laminate substrate was pressed onto the side of the YIG pillars with tape, which affixed the YIG pillars tightly and precisely onto the copper clad laminate substrate. The acrylic mold was removed carefully to prevent extrusion and damage to the YIG pillars.

Each YIG pillar was properly magnetized by aligning the NdFeB magnet used to apply the magnetic field precisely underneath each YIG pillar. To this end, a 2-mm-thick acrylic sheet was first covered with a double-sided tape to form the substrate. Then, we fixed a 5-mm-thick acrylic sheet including 10 × 10 perforations with a radius of 5 mm and a lattice constant of 17.9 mm onto a 2-mm-thick acrylic substrate. Finally, with respect to the designated direction of the magnetic field, we placed the NdFeB magnets into the holes of the 5-mm-thick acrylic sheet, achieving a good alignment with the array of the YIG pillars (Figure 1b).

**Characterization Setup.** We measured the photonic band structure, transmission spectra, and near-field distribution of the active metamaterial sample using two dipole antennas as the transmitter and receiver. Both antennas were inserted into the waveguide via holes drilled using an ultrahigh-precision CNC machine and they were connected to a vector network analyzer (Rohde & Schwarz ZNB 20) to measure the S parameters. Prior to measurement, 3.5-mm 85052D through-open-short-load calibrations were performed. As a result, the measured S parameters included only the insertion loss of the tapered waveguides and the metamaterial.

## Acknowledgements

The authors appreciate the discussions with Hengbin Cheng from the Institute of Physics, Chinese Academy of Science, and their assistance. The authors are also grateful for the support received from the Ministry of Science and Technology of the People's Republic of China (MOST) (Grant No. 2018YFE0109200, 2021YFA1401000, and 2021YFB2801600), National Natural Science Foundation of China (NSFC) (Grant Nos. 51972044, 52021001, and 62075114), Sichuan Provincial Science and Technology Department (Grant No. 2019YFH0154), the Fundamental Research Funds for the Central Universities (Grant No. ZYGX2020J005), the Beijing Natural Science Foundation (Grant No. 4212050), and the Zhuhai Industry-University Research Collaboration Project (ZH22017001210108PWC). This work was supported by the Center of High-Performance Computing, Tsinghua University.